\documentclass[12pt,english]{elsarticle}
\usepackage[T1]{fontenc}
\usepackage[latin9]{inputenc}
\usepackage{verbatim}
\usepackage{amsmath}
\usepackage{amsthm}
\usepackage{amssymb}
\usepackage{graphicx}
\usepackage[a4paper]{geometry}
\makeatletter
\theoremstyle{plain}
\newtheorem{lem}{\protect\lemmaname}
\theoremstyle{plain}
\newtheorem{thm}{\protect\theoremname}
\theoremstyle{plain}
\newtheorem{cor}{\protect\corollaryname}

\@ifundefined{date}{}{\date{}}
\usepackage{multicol}

\newcommand{\vs}{\vspace}

\makeatletter
\def\ps@pprintTitle{
 \let\@oddhead\@empty
 \let\@evenhead\@empty
 \let\@oddfoot\@empty
 \let\@evenfoot\@empty
}
\makeatother

\makeatother

\usepackage{babel}
\providecommand{\corollaryname}{Corollary}
\providecommand{\lemmaname}{Lemma}
\providecommand{\theoremname}{Theorem}

\begin{document}
\begin{frontmatter}
\title{On the Scientific Method: The Role of Hypotheses and Involved Mathematics}
\author{Mario Milanese$^{a}$, Carlo Novara$^{a}$, Michele Taragna$^{a}$}
\address{$^{a}$Politecnico di Torino \\ \vspace{1em}\centerline{Email: mario.milanese@formerfaculty.polito.it, carlo.novara@polito.it, michele.taragna@polito.it}}
\begin{abstract}
\vs{0mm}The paper investigates the role of data, hypotheses and mathematical
methods that can be used in the discovery of the law $y=\mathit{f_{o}}(u)$,
relating variables $u$ and $y$ of a physical phenomenon, making
use of experimental measurements of such variables. Since the exact
knowledge of $\mathit{f_{o}}$ cannot be expected, the problem of
deriving approximate functions $\widehat{f}$giving small error $\mathit{f_{o}}-\widehat{f}$
measured by some function norm is discussed. The main contributions
of the paper are summarized as follows. 

At first, it is proven that deriving a reliable approximate $\widehat{f}$,
i.e., having a finite error, is not possible using measured data only.
Thus, for deriving reliable approximate $\mathit{\widehat{f}}$, hypotheses
on the function $f_{o}$ and on the disturbances corrupting the measurements
must be introduced.

Second, necessary and sufficient conditions for deriving reliable
$\mathit{\widehat{f}}$ are provided. If such conditions are satisfied,
suitable accuracy properties of reliable approximate $\mathit{\widehat{f}}$
can be defined (e.g., ordering relationship among them, optimality,\dots ),
called theoretical properties.

Third, it is shown that it is not possible to verify the conditions
necessary for deriving reliable $\mathit{\widehat{f}}$, but that
it is possible to verify that hypotheses on $\mathit{f_{o}}$ and
on the disturbances are falsified by experimental measurements, showing
that no function and disturbances satisfying the given hypotheses
exist, able to reproduce the measurements (this is called falsification
property).

The above properties are then discussed for hypotheses belonging to
the following classes: Parametric Probabilistic, where $\mathit{f_{o}}$
is assumed to be a function depending on a vector $p$ and the disturbances
are assumed to be stochastic variables; Set Membership class, where
$\mathit{f_{o}}$ is assumed to be a bounded smooth function and the
disturbances are assumed to be bounded variables; Parametric Set Membership
class, able to integrate Parametric Probabilistic hypotheses with
Set Membership hypotheses.

\end{abstract}
\end{frontmatter}

\section{Introduction\label{sec:Introduction}}

The revolutionary studies started in the 17th Century on scientific
methods for investigating physical phenomena, based on doing experiments,
measurements and verifications, has been fundamental in human civilization.

A huge literature on this topic has been since then developed, and
different kinds of investigation have been considered. In this paper
we consider the wide class of scientific problems where a physical
phenomenon P is investigated, making use of a set of measured data
$D_{o}=\left\{ \widetilde{u}_{k},\widetilde{y}_{k};k=1,2,\ldots,M\right\} $
of two types of variables: $u\in R^{n_{u}}$, whose values can be
selected by experimenters, and all the other measured variables $y\in R^{n_{y}}$.
The aim of scientific investigation is to ``discover'' if there
exists some law $f_{o}$, that relates the measured variables $y$
and $u$ as $y=f_{o}\left(u\right)$. 

Since the pioneering works of Descartes, Galileo, Newton, Hume, and
until to present time, many methodologies have been proposed for achieving
this aim, motivated by different emphases on inductive versus hypothetical-deductive
reasoning, and giving rise to large and unending debates among different
schools, see, e.g., \cite{Popp35,Jeff56,Kuhn62,Carn67,NoSa00,SuCo19}.

The aim of the paper is to formulate such class of scientific investigations
as a problem of ``inference making from data'', see \cite{MiNo11,TrWW88n}
and the references therein, and discuss how this formulation may give
interesting insights of the role of data, hypotheses and mathematical
methods that can be used in the discovery of the laws of specific
physical phenomena and on the evaluation of their achievable accuracy
properties.

\section{The inference making from data formulation\label{sec:Inference-making}}

The function $f_{o}\in F=\left\{ f:R^{n_{u}}\rightarrow R^{n_{y}}\right\} $
to be ``discovered'' is not known, but a set of measurements of
variables $u$ and $y$ has been performed and collected in the data
set $D_{o}=\left\{ \widetilde{u}_{k},\widetilde{y}_{k};k=1,\ldots,M\right\} $,
where $\widetilde{u}_{k}=u_{k}+\mathit{du}_{k}$, $\widetilde{y}_{k}=y_{k}+\mathit{dy}_{k}$,
and $\mathit{du}_{k}$, $\mathit{dy}_{k}$ are unknown disturbances
affecting the measurement of $u$ and $\mathit{y}$. The information
on $f_{o}$ provided by data set $D_{o}$ is described by the following
set of equations:
\begin{equation}
\widetilde{y}_{k}=f_{o}\left(\widetilde{u}_{k}\right)+d_{k},\quad k=1,\ldots,M\label{eq:measurements}
\end{equation}
where $d_{k}=f_{o}(u_{k})-f_{o}(\widetilde{u}_{k})-\mathit{dy}_{k}$
is a disturbance accounting for the disturbances in $u$ and $y$
measurements.

It cannot be expected that such information may be sufficient to know
$f_{o}$ exactly. Thus, the problem is to find methods able to derive
functions $\widehat{f}\in F$ to be used as ``suitable'' approximations
of $f_{o}$. For the sake of exposition simplicity, in the following
it is considered that $n_{y}=1$ and that $f_{o}\in C^{0}$, where
$C^{0}$ is the class of all continuous functions. Any such method
is represented as an operator $\widehat{\phi}:D_{o}\rightarrow C^{0}$,
called here estimator, that, making use of data set $D_{o}$, derives
a function $\widehat{f}$, called estimate of $f_{o}$. Let $\Phi\left(D_{o}\right)$
be the set of all possible estimators $\phi:D_{o}\rightarrow C^{0}$.
A key point is to define how to measure the approximation accuracy
of $\widehat{f}$ derived by given estimator $\widehat{\phi}\in\Phi\left(D_{o}\right)$. 

The accuracy of $\widehat{f}$ has to be measured not only evaluating
that all errors $e_{k}=f_{o}\left(\widetilde{u}_{k}\right)-\widehat{f}\left(\widetilde{u}_{k}\right)$,
called fitting errors, are ``small'', but also evaluating that errors
$e_{p}\left(\widetilde{u}\right)=f_{o}\left(\widetilde{u}\right)-\widehat{f}\left(\widetilde{u}\right)$
for $\widetilde{u}\neq\widetilde{u}_{k}$, called predicted errors,
are ``small''. This property is often referred to in the literature
as ``generalization'' property. 

$L_{q}(U)$ norms $\left\Vert f\right\Vert _{q}^{U}$, $q=1,\ldots,\infty$,
allow formal definitions of this property, evaluating the accuracy
of predicted errors $e_{p}\left(\widetilde{u}\right)=f_{o}\left(\widetilde{u}\right)-\widehat{f}\left(\widetilde{u}\right)$,
$\forall\widetilde{u}\in U$, where $U$ is the smallest box containing
$D_{o}$, defined as: 
\begin{equation}
U=\left\{ \widetilde{u}\in R^{n_{u}}:\widetilde{u}\left(i\right)\in\left[\underline{u}\left(i\right),\overline{u}\left(i\right)\right],i=\left[1,\ldots,n_{u}\right]\right\} ,\;\underline{u}\left(i\right)=\underset{k}{\mathit{\min}}\widetilde{u}_{k}\left(i\right),\;\overline{u}\left(i\right)=\underset{k}{\mathit{\max}}\widetilde{u}_{k}\left(i\right).\label{eq:set_U}
\end{equation}
For example, if $\left\Vert f_{o}-\widehat{f}\right\Vert _{\infty}^{U}=\epsilon$,
then $\underset{\widetilde{u}\in U}{\max}\left|e_{p}\left(\widetilde{u}\right)\right|\leq\epsilon$.
Thus, not only all the $M$ fitting errors $e_{k}$ have absolute
values bounded by $\epsilon$, but this is true also for all predicted
errors $e_{p}\left(\widetilde{u}\right),\widetilde{u}\in U$. If $\left\Vert f_{o}-\widehat{f}\right\Vert _{2}^{U}=\epsilon$,
then $\left\Vert f_{o}-\widehat{f}\right\Vert _{2}^{U}=\sqrt[2]{\int_{U}\left(e_{p}\left(\widetilde{u}\right)\right)^{2}\mathit{d\widetilde{u}}}=\epsilon$,
i.e., $\epsilon$ is the root of all square errors $e_{p}\left(\widetilde{u}\right)$
integrated on the set $U$.

Thus, it is relevant to investigate the existence of some estimator
$\widehat{\phi}\in\Phi\left(D_{o}\right)$ that derives approximate
function $\widehat{f}$ having finite approximation error $\left\Vert f_{o}-\widehat{f}\right\Vert _{q}^{U}$.
Estimators having this property are called ``reliable''. For example,
if $\widehat{\phi}\in\Phi\left(D_{o}\right)$ is not reliable with
$q=\infty$, then, for some $\widetilde{u}\in U$, the predicted errors
$e_{p}\left(\widetilde{u}\right)=f_{o}\left(\widetilde{u}\right)-\widehat{f}\left(\widetilde{u}\right)$
may be arbitrarily large.

\bigskip{}

\textbf{Remark (Dynamical phenomena)}: It may seem that as above formulated,
the investigation of ``static'' phenomena only may be considered,
where variables $u$ and $y$ are measured when they have constant
values. Indeed, the formulation allows to consider causal dynamic
phenomena, where variables $u$ and $y$ are not constant in time
and measured at times $j\cdot\Delta t$ for given sampling time $\Delta t$.\\
Dynamic phenomena are characterized by the fact that the value of
$y$ at time $j\cdot\Delta t$, $y_{j}$ for short, depends not only
on $u_{j}$, but also on past values of $u$ and $y$, according to
a regression function $y_{j}=f_{o}\left(r_{j}\right)$, where $r_{j}=\left[y_{j-1},\ldots,y_{j-m},u_{j}\left(1\right),\ldots,u_{j-m}\left(1\right),\ldots,u_{j}\left(n_{u}\right),\ldots,\right.$
$\left.u_{j-m}\left(n_{u}\right)\right]$, for some value of $m$.
Thus, in the investigation of dynamic systems, the problem is to approximate
the function $f_{o}$ having a set of noise corrupted measurements
$D_{o}=\left\{ \widetilde{r}_{j},\widetilde{y}_{j};j=1,2,\ldots,M\right\} $.
Thus, the results reported in the paper hold also for nonlinear dynamic
systems.\hfill$\blacksquare$

\bigskip{}

A preliminary lemma is stated,\,related to the case of exact measurements,\,where
the data set is\,$D_{\!\mathit{ex}}\!\!=\!\!\left\{ u_{k},\!y_{k};k\!=\!1,\ldots,M\right\} $.
\begin{lem}
\label{lem:Lemma1}No reliable estimators $\widehat{\phi}\in\Phi\left(D_{\mathit{ex}}\right)$
exist, with $q=1,\ldots,\infty$, whatever large the number $M$ of
data set $D_{\mathit{ex}}$ is.
\end{lem}
\bigskip{}

\textbf{Proof.} Let us define the set $\mathit{FFS}\left(D_{\mathit{ex}}\right)=\left\{ f\in C^{0}:y_{k}=f\left(u_{k}\right),k=1,\ldots,M\right\} $.
This is the set of all continuous functions that could have generated
the data set $D_{\mathit{ex}}$. Thus, the information (\ref{eq:measurements})\textbf{
}provided by data set $D_{\mathit{ex}}$ on the unknown $f_{o}$ is
that $f_{o}\in\mathit{FFS}\left(D_{\mathit{ex}}\right)$. 

Let $\Phi\left(D_{\mathit{ex}}\right)$ be the set of all possible
estimators $\phi:D_{\mathit{ex}}\rightarrow C^{0}$. For given $\widehat{\phi}\in\Phi\left(D_{\mathit{ex}}\right)$,
the value $E_{q}\left(\widehat{\phi}\right)\triangleq\underset{f\in\mathit{FFS}\left(D_{\mathit{ex}}\right)}{\sup}\left\Vert f-\widehat{f}\right\Vert _{q}^{U}$
is the smallest value that can be guaranteed on $\left\Vert f-\widehat{f}\right\Vert _{q}^{U}$
on the base of the information on $f_{o}$ provided by data set $D_{o}$.
Now it is proved that:
\begin{equation}
E_{q}\left(\widehat{\phi}\right)=\infty,\quad\forall\widehat{\phi}\in\Phi\left(D_{\mathit{ex}}\right).\label{eq:error}
\end{equation}
Let $\widetilde{u}\in U$ be such that $\widetilde{u}\neq u_{k},\forall k.$
Let $f_{I}^{b}$ be a continuous function obtained by any interpolation
method with nodes $f_{I}^{b}\left(\widetilde{u}\right)=b$, $f_{I}^{b}\left(u_{k}\right)=y_{k},k=1,\ldots,M$.
Since $f_{I}^{b}\in\mathit{FFS}\left(D_{\mathit{ex}}\right),\forall b$,
then, $E_{q}\left(\widehat{\phi}\right)\geq\left\Vert f_{I}^{b}-\widehat{f}\right\Vert _{q}^{U},\forall b$. 

If $q=\infty$, $\left\Vert f_{I}^{b}-\widehat{f}\right\Vert _{\infty}^{U}=\underset{u\in U}{\max}\left|f_{I}^{b}\left(u\right)-\widehat{f}\left(u\right)\right|$,
so that $\left\Vert f_{I}^{b}-\widehat{f}\right\Vert _{\infty}^{U}\geq\left|f_{I}^{b}\left(\widetilde{u}\right)-\widehat{f}\left(\widetilde{u}\right)\right|$.
Thus, $E_{\infty}\left(\widehat{\phi}\right)\geq\underset{b\rightarrow\infty}{\lim}\left\Vert f_{I}^{b}-\widehat{f}\right\Vert _{\infty}^{U}\geq\underset{b\rightarrow\infty}{\lim}\left|f_{I}^{b}\left(\widetilde{u}\right)-\widehat{f}\left(\widetilde{u}\right)\right|$.
Since $\underset{b\rightarrow\infty}{\lim}\left|f_{I}^{b}\left(\widetilde{u}\right)-\widehat{f}\left(\widetilde{u}\right)\right|\rightarrow\infty$,
(\ref{eq:error}) is proved for $q=\infty$.

If $q\in[1,\infty)$, $\left\Vert f_{I}^{b}-\widehat{f}\right\Vert _{q}^{U}=\left[\int_{U}\left|f_{I}^{b}\left(u\right)-\widehat{f}\left(u\right)\right|^{q}\mathit{du}\right]^{1/q}$
and $\left\Vert f_{I}^{b}-\widehat{f}\right\Vert _{q}^{U}\geq\left\Vert f_{I}^{b}\right\Vert _{q}^{U}-\left\Vert \widehat{f}\right\Vert _{q}^{U}$.
Since $\underset{b\rightarrow\infty}{\lim}\left|f_{I}^{b}\left(u\right)\right|\rightarrow\infty$
and $f_{I}^{b}$ is continuous, $\underset{b\rightarrow\infty}{\lim}\left[\int_{U}\left|f_{I}^{b}\left(u\right)\right|^{q}\mathit{du}\right]^{1/q}\rightarrow\infty$,
thus concluding the proof of (\ref{eq:error}) for any $q\in[1,\infty]$.\hfill$\blacksquare$

\bigskip{}

From Lemma~\ref{lem:Lemma1} the following result for the case of
corrupted measurements trivially follows. 
\begin{thm}
\label{thm:Th1}No reliable estimators $\widehat{\phi}\in\Phi\left(D_{o}\right)$
exist, with $q=1,\ldots,\infty$, whatever large the number $M$ of
data set $D_{o}$ is.
\end{thm}
\bigskip{}

Thus, reliable estimators could be derived only by making suitable
hypotheses on the unknown function $f_{o}\in F^{a}$ and on unknown
disturbance variable $d\in D^{a}$. Note that from Lemma~\ref{lem:Lemma1}
it follows that $F^{a}\subset C^{0}$ is a necessary condition for
deriving reliable estimators. Thus, in the rest of the paper only
hypotheses $F^{a}\subset C^{0}$ are considered.

\bigskip{}
The most widely hypotheses used in the scientific literature belong
to the following class, called in this paper Parametric Probabilistic
(PP) hypotheses, see, e.g., \cite{Kolm41,Kush64,Stra68,Jazw70n,Gelb74n,Mayb82,SZLB95,Vapn00}
and the references therein:
\begin{itemize}
\item Parametric Probabilistic (PP) hypotheses:
\begin{itemize}
\item $f_{o}\in F_{p}^{a}$, set of continuous functions $f_{p}\left(u\right)$
depending on parameter vector $p\in R^{n_{p}}$ 
\item $d\in D_{s}^{a}$, set of stochastic variables with given probability
density function
\end{itemize}
\end{itemize}
Estimators $\widehat{\phi}\in\Phi(D_{o},F_{p}^{a},D_{s}^{a})$, making
use of PP hypotheses, are called PP estimators. 

\bigskip{}

In the last few decades, hypotheses belonging to the following class,
called Set Membership (SM) hypotheses, have been used, see, e.g.,
\cite{Schw73,MiRi77,MiBe82,MNPW96,MiNo04,MiNo07,MoKC09,NoRM13} and
the references therein:
\begin{itemize}
\item Set Membership (SM) hypotheses:
\begin{itemize}
\item $f_{o}\in F_{\gamma}^{a}$, set of ``smooth'' continuous functions
$f_{\gamma}\left(u\right)$ with smoothness parameter $\gamma\in R$ 
\item $d\in D_{\varepsilon}^{a}$, set of disturbances $d$ such that $\left|d\right|\leq\varepsilon$ 
\end{itemize}
\end{itemize}
Estimators $\widehat{\phi}\in\Phi(D_{o},F_{\gamma}^{a},D_{\varepsilon}^{a})$,
making use of SM hypotheses, are called SM estimators.

\bigskip{}

In the next two sections, two basic properties of estimators are investigated
and it is discussed how they apply to PP and SM estimators.

\section{Theoretical Accuracy Properties of Estimators\label{sec:Theoretical-Accuracy-Properties}}

Let $\Phi(D_{o},F^{a},D^{a})$ be the set of all estimators $\phi:(D_{o},F^{a},D^{a})\rightarrow C^{0}$
that derive functions $f\in C^{0}$, to be used as approximation of
$f_{o}$, using information (\ref{eq:measurements}) provided by data
$D_{o}$, and hypotheses $f_{o}\in F^{a}$, $d\in D^{a}$. Function
$\widehat{f}$ derived by estimator $\widehat{\phi}\in\Phi(D_{o},F^{a},D^{a})$,
called in this paper estimate of $f_{o}$, in the literature is called
with many different terms, e.g., model, theory, digital twin. 

The following theorem gives necessary and sufficient conditions for
the existence of reliable estimators $\widehat{\phi}\in\Phi(D_{o},F^{a},D^{a})$.

Let set $FFS\left(D_{o},F^{a},D^{a}\right)$, called Feasible Function
Set, be defined as: 
\begin{equation}
FFS\left(D_{o},F^{a},D^{a}\right)=\left\{ f\in F^{a}:\widetilde{y}_{k}=f\left(\widetilde{u}_{k}\right)+d_{k},k=1,\ldots,M,\;d\in D^{a}\right\} \label{eq:FFS}
\end{equation}
This is the set of all functions $f\in F^{a}$ that can generate the
measurements $\widetilde{y}_{k}$ from measurements of $\widetilde{u}_{k}$,
with disturbances $d\in D^{a}$. Let $\overline{FFS}\left(D_{o},F^{a},D^{a}\right)$
be the closure of $FFS\left(D_{o},F^{a},D^{a}\right)$ with respect
to $L_{\infty}(U)$ norm.

\bigskip{}

\begin{thm}
\label{thm:Th2}Reliable Estimators $\widehat{\phi}\in\Phi(D_{o},F^{a},D^{a})$
exist if and only if $FFS\left(D_{o},F^{a},D^{a}\right)$ is not empty
and bounded and $f_{o}\in\overline{FFS}\left(D_{o},F^{a},D^{a}\right)$.
\end{thm}
\bigskip{}

\textbf{Proof}. Let $\widehat{f}$ be derived by an estimator $\widehat{\phi}\in\Phi(D_{o},F^{a},D^{a})$
making use of hypotheses $F^{a},D^{a}$ such that the two conditions
$FFS\left(D_{o},F^{a},D^{a}\right)$ is not empty and bounded, and
$f_{o}\in\overline{FFS}\left(D_{o},F^{a},D^{a}\right)$ hold true.
Norm triangle inequality shows that $\left\Vert f_{o}-\widehat{f}\right\Vert _{q}^{U}\leq\left\Vert f_{o}\right\Vert _{q}^{U}+\left\Vert \widehat{f}\right\Vert _{q}^{U}$,
where $\left\Vert \widehat{f}\right\Vert _{q}^{U}$ is a finite value,
since $\widehat{f}\in C^{0}$ and $U$ is a compact set. If $f_{o}\in\overline{FFS}\left(D_{o},F^{a},D^{a}\right)$,
the following bound holds:
\begin{equation}
\left\Vert f_{o}-\widehat{f}\right\Vert _{q}^{U}\leq\underset{f\in\overline{FFS}\left(D_{o},F^{a},D^{a}\right)}{\sup}\left\Vert f\right\Vert _{q}^{U}+\left\Vert \widehat{f}\right\Vert _{q}^{U}\label{eq:upper_bound}
\end{equation}
If $FFS\left(D_{o},F^{a},D^{a}\right)$ is bounded, $\overline{FFS}\left(D_{o},F^{a},D^{a}\right)$
also is bounded and then the RHS of (\ref{eq:upper_bound}) is finite,
proving the if condition. 

The necessity of not emptiness of $FFS\left(D_{o},F^{a},D^{a}\right)$
follows from the fact that if \\ $FFS\left(D_{o},F^{a},D^{a}\right)$
is empty, no functions $f\in F^{a}$ exist that can generate the measurements
$\widetilde{y}_{k}$ from measurements of $\widetilde{u}_{k}$, with
disturbances $d\in D^{a}$. Thus, it is not known what function $f$
estimators $\widehat{\phi}\in\Phi(D_{o},F^{a},D^{a})$ should try
to approximate.

The necessity of condition of $FFS\left(D_{o},F^{a},D^{a}\right)$
boundedness is now proved. From norm reversed triangle inequality
the following inequality holds:
\begin{equation}
\sup_{f\in\mathit{FFS}\left(D_{o},F^{a},D^{a}\right)}\left\Vert f-\widehat{f}\right\Vert _{q}^{U}\geq\underset{f\in\mathit{FFS}\left(D_{o},F^{a},D^{a}\right)}{\sup}\left|\left\Vert f\right\Vert _{q}^{U}-\left\Vert \widehat{f}\right\Vert _{q}^{U}\right|\label{eq:lower_bound_1}
\end{equation}
If $FFS\left(D_{o},F^{a},D^{a}\right)$ is not bounded, the RHS of
(\ref{eq:lower_bound_1}) can be arbitrarily large.

The necessity of condition $f_{o}\in\overline{\mathit{FFS}}\left(D_{o},F^{a},D^{a}\right)$
is now proved. From norm reversed triangle inequality, the following
inequality holds:
\begin{equation}
\left\Vert f_{o}-\widehat{f}\right\Vert _{q}^{U}\geq\left|\left\Vert f_{o}\right\Vert _{q}^{U}-\left\Vert \widehat{f}\right\Vert _{q}^{U}\right|\label{eq:lower_bound_2}
\end{equation}
where $\left\Vert \widehat{f}\right\Vert _{q}^{U}$ is a finite value.
If $f_{o}\notin\overline{FFS}\left(D_{o},F^{a},D^{a}\right)$, the
RHS of (\ref{eq:lower_bound_2}) can be arbitrarily large even if
$FFS\left(D_{o},F^{a},D^{a}\right)$ is bounded, thus concluding the
proof of the only if condition.\hfill$\blacksquare$

\bigskip{}

The two conditions $FFS\left(D_{o},F^{a},D^{a}\right)$ is not empty
and bounded and \\ $f_{o}\in\overline{FFS}\left(D_{o},F^{a},D^{a}\right)$
are called Basic Accuracy (BA) conditions and, separately, BA1 and
BA2 conditions. Theorem~\ref{thm:Th2} states that, under the BA
conditions, reliable estimators $\phi\in\Phi\left(D_{o},F^{a},D^{a}\right)$
exist. Indeed, further relevant properties of estimators $\phi\in\Phi\left(D_{o},F^{a},D^{a}\right)$
can be derived, as shown by the following corollary of Theorem~\ref{thm:Th2}.
\begin{cor}
\label{cor:Corollary1}If a reliable estimator $\phi\in\Phi\left(D_{o},F^{a},D^{a}\right)$
exists, then any $\widehat{\phi}\in\Phi\left(D_{o},F^{a},D^{a}\right)$
is reliable, with finite approximation error bound given as:
\begin{equation}
\left\Vert f_{o}-\widehat{f}\right\Vert _{q}^{U}\leq\sup_{f\in\overline{FFS}\left(D_{o},F^{a},D^{a}\right)}\left\Vert f-\widehat{f}\right\Vert _{q}^{U}\triangleq E_{q}\left(\widehat{\phi}\right)\label{eq:corollary}
\end{equation}
\end{cor}
\textbf{Proof.} Since it is assumed that a reliable estimator $\phi\in\Phi\left(D_{o},F^{a},D^{a}\right)$
exists, from Theorem~\ref{thm:Th2} it follows that $FFS\left(D_{o},F^{a},D^{a}\right)$
is not empty and bounded and that $f_{o}\in\overline{FFS}\left(D_{o},F^{a},D^{a}\right)$,
thus proving the inequality (\ref{eq:corollary}). Indeed, $\overline{FFS}\left(D_{o},F^{a},D^{a}\right)$
is bounded, being the closure of a bounded set, thus proving that
the RHS of (\ref{eq:corollary}) is finite.\\ \hspace*{10cm}\hfill$\blacksquare$

Quantity $E_{q}\left(\widehat{\phi}\right)$ is called approximation
error of $\widehat{\phi}$ and can be used as index of estimators
reliability, thus inducing the following ordering relationship among
all reliable estimators $\phi\in\Phi\left(D_{o},F^{a},D^{a}\right)$:
\begin{equation}
\widehat{\phi}\prec\breve{\phi}\leftrightarrow E_{q}\left(\widehat{\phi}\right)<E_{q}\left(\breve{\phi}\right),\quad\forall\widehat{\phi},\breve{\phi}\in\Phi\left(D_{o},F^{a},D^{a}\right)
\end{equation}
This ordering relationship allows the definition of the following
optimality concepts. 

Reliable estimator $\phi^{\ast}\in\Phi\left(D_{o},F^{a},D^{a}\right)$
is called optimal if: 
\begin{equation}
E_{q}\left(\phi^{\ast}\right)\leq E_{q}\left(\widehat{\phi}\right),\quad\forall\widehat{\phi}\in\Phi\left(D_{o},F^{a},D^{a}\right)
\end{equation}

\bigskip{}

It is now discussed if the following properties can be derived by
PP or SM reliable estimators.
\begin{enumerate}
\item Derivation of optimal estimators $\phi^{\ast}$ 
\item Computation of errors $E_{q}\left(\phi^{\ast}\right)$ 
\item If derivation of optimal estimator is not known, or too computationally
demanding, derivation of an estimator $\widehat{\phi}\in\Phi\left(D_{o},F^{a},D^{a}\right)$
having known suboptimality level $\alpha=E_{q}\left(\widehat{\phi}\right)/E_{q}\left(\phi^{\ast}\right)$
.
\end{enumerate}
\bigskip{}

These accuracy properties are called ``theoretical'', since are
deduced assuming that the BA conditions of Theorem~\ref{thm:Th2}
hold true.

\subsection{Theoretical accuracy properties of PP estimators\label{subsec:Theoretical-accuracy-of-PP}}

Two classes of parametric hypotheses $F_{p}^{a}$ are used in the
literature. The first class of hypotheses, indicated here as ``physical''
hypotheses, is obtained by deriving suitable mathematical equations,
based on the laws governing the investigated phenomenon, and depending
on unknown parameters having relevant physical interpretation (e.g.,
body masses for mechanical systems, damping coefficients for motion
in fluids, ...). Since the beginning of the scientific revolution,
physical hypotheses $F_{p}^{a}$ have been the basis of scientific
investigations. In the last few decades, many machine learning methods
have been developed (e.g., Neural Nets, Deep Learning, Support Vector
Regression, ...) using ``black-box'' hypotheses $F_{p}^{a}$, in
order to deal with phenomena for which not sufficiently reliable governing
laws are known. Black-box hypotheses define $F_{p}^{a}$ as combination
of nonlinear basis functions (wavelets, sigmoids, radial basis functions,
kernels, ...), defined by different machine learning methods.

Optimal estimators have been derived under very specific PP hypotheses,
e.g., $F_{p}^{a}=F_{p}^{L}$ , set of linear functions or linear combination
of given nonlinear functions $f_{i}\left(u\right),i=1:n_{p}$, $D_{s}^{a}$
is a set of stochastic variables having Gaussian or few other probability
density functions, and approximation error is measured by $L_{2}$
norm. However, in the case that $F_{p}^{a}$ is a set of ``complex''
nonlinear functions, i.e., it is not a linear combination of a finite
number of given nonlinear functions, finding optimal estimators is
computationally intractable, see, e.g., \cite{Kush64,Stra68}. Many
methods have been proposed to derive computationally tractable estimators
$\widehat{\phi}\in\Phi\left(D_{o},F_{p}^{a},D_{s}^{a}\right)$, based
on choosing as estimate of $f_{o}$ the function $f_{\widehat{p}}\in F_{p}^{a}$,
where $\widehat{p}$ minimizes a given cost function $J\left(\widetilde{Y},\widehat{Y}\left(p\right)\right)$,
where $\widetilde{Y}=\left[\widetilde{y}_{1},\ldots,\widetilde{y}_{M}\right]$,
$\widehat{Y}\left(p\right)=\left[f_{p}\left(\widetilde{u}_{1}\right),\ldots,f_{p}\left(\widetilde{u}_{M}\right)\right]$. 

For all such computationally tractable methods, it is in general hard
to known how far the derived estimator $\widehat{\phi}$ is from optimality.
Even reliable evaluation of error $E_{q}\left(\widehat{\phi}\right)$
may not be achieved, since in general not sufficiently tight lower
and upper bounds of $E_{q}\left(\widehat{\phi}\right)$ can be derived. 

\subsection{Theoretical accuracy properties of SM estimators\label{subsec:Theoretical-accuracy-of-SM}}

Some results are here reported, related to the case that $F_{\gamma}^{a}$
is the set of continuous differentiable nonlinear functions on set
$U$, having $\left\Vert f^{\prime}(u)\right\Vert _{2}\leq\gamma,\:\forall u\in U$,
where $f^{\prime}(u)\in R^{n_{u}}$ is the gradient of $f(u)$ and
$\left\Vert \cdot\right\Vert _{2}$ is the Euclidean norm. Very similar
properties are obtained assuming that $F_{\gamma}^{a}$ is the set
of Lipschitz continuous functions with Lipschitz constant bounded
by $\gamma$ on $U$. 

For any $\gamma,\varepsilon$ values such that $\mathit{FFS}\left(D_{o},F_{\gamma}^{a},D_{\varepsilon}^{a}\right)$
is not empty and bounded, let functions $\underline{f}$, $\overline{f}$
and $f^{c}$ be defined as: 
\begin{equation}
\underline{f}\left(u\right)=\max_{k=1,\ldots,M}\left[\left(\widetilde{y}_{k}-\varepsilon\right)-\gamma\left\Vert u-\widetilde{u}_{k}\right\Vert _{2}\right],\quad u\in U\label{eq:f_lower}
\end{equation}
\begin{equation}
\overline{f}\left(u\right)=\min_{k=1,\ldots,M}\left[\left(\widetilde{y}_{k}+\varepsilon\right)+\gamma\left\Vert u-\widetilde{u}_{k}\right\Vert _{2}\right],\quad u\in U\label{eq:f_upper}
\end{equation}
\begin{equation}
f^{c}\left(u\right)=\left[\underline{f}\left(u\right)+\overline{f}\left(u\right)\right]/2\label{eq:f_center}
\end{equation}

Let $\phi^{c}\in\Phi\left(D_{o},F_{\gamma}^{a},D_{\varepsilon}^{a}\right)$
be the estimator deriving as estimate function $f^{c}$. Then, the
following theoretical results hold, see \cite[Theorems~2,7]{MiNo04}:
\begin{enumerate}
\item $\phi^{c}$ is an optimal estimator for any $L_{q}\left(U\right)$
norm, $q=1,2,\ldots,\infty$
\item $E_{q}\left(\phi^{c}\right)=\left\Vert \overline{f}-\underline{f}\right\Vert _{q}^{U}/2$
\item $\underline{f}\left(\widetilde{u}\right)\leq f_{o}\left(\widetilde{u}\right)\leq\overline{f}\left(\widetilde{u}\right),\quad\forall\widetilde{u}\in U$
\end{enumerate}
\bigskip{}

The error $E_{q}\left(\phi^{c}\right)$ can be evaluated computing
the $L_{q}\left(U\right)$ norm of the known function $\overline{f}-\underline{f}$.
It must be noted that the complexity in evaluating $L_{1}\left(U\right)$
and $L_{2}\left(U\right)$ norms growths exponentially with $n_{u}$,
resulting very high for $n_{u}$ greater than $10-15$. In the case
that $L_{\infty}\left(U\right)$ norm is considered, of large interest
in many investigations, a method has been proposed in \cite{MiNo07},
whose computational complexity in computing $E_{\infty}\left(\phi^{c}\right)$
is $\mathcal{O}\left(n_{u}^{2}\right)$, thus allowing its evaluation
for quite large values of $n_{u}$.

The computing time and memory occupation for achieving such optimality
properties are now discussed. The time for evaluating $f^{c}\left(u\right)$
grows linearly with $M$. On standard personal computers, the evaluation
of (\ref{eq:f_center}) with $M=10^{6}$ has computing times of the
order of 1 ms. The memory occupation is $M\cdot\left(n_{u}+1\right)$
bytes, growing linearly with $M$. 

In case that one or both computational requirements cannot be satisfied
by the available electronic equipment, an estimator $\phi_{P}\in\Phi\left(D_{o},F_{\gamma}^{a},D_{\varepsilon}^{a}\right)$,
able to trade-off between computational requirements and estimation
accuracy, can be derived as follows. The estimator $\phi_{P}$ derives
$f_{\widehat{p}}$ within a parametric set of functions $f_{p}\in F_{p}$,
where $\widehat{p}$ minimizes the cost function $J\left(p\right)=\left\Vert \widetilde{Y}-\widehat{Y}\left(p\right)\right\Vert _{\infty}$,
$\widetilde{Y}=\left[\widetilde{y}_{1},\ldots,\widetilde{y}_{M}\right]$,
$\widehat{Y}\left(p\right)=\left[f_{p}\left(\widetilde{u}_{1}\right),\ldots,f_{p}\left(\widetilde{u}_{M}\right)\right]$,
with constraint $f_{\widehat{p}}\in F_{\gamma}^{a}$. If $J\left(\widehat{p}\right)\leq\varepsilon$,
it results that $f_{\widehat{p}}\in\mathit{FFS}\left(D_{o},F_{\gamma}^{a},D_{\varepsilon}^{a}\right)$,
which implies that $E_{q}\left(\phi_{P}\right)/E_{q}\left(\phi^{c}\right)=\alpha\leq2$.

Thus, the operator $\phi_{P}\in\Phi\left(D_{o},F_{\gamma}^{a},D_{\varepsilon}^{a}\right)$
has the following properties: 

a) computing time and memory occupation essentially independent from
the number of data 

b) suboptimality level $\alpha=E_{q}\left(\phi_{P}\right)/E_{q}\left(\phi^{\ast}\right)\leq2$

\bigskip{}

This discussion can be summarized as follows: 
\begin{itemize}
\item SM estimators, achieving the above defined ``theoretical'' accuracy
properties, can be derived for general nonlinear hypotheses on function
$f_{o}$, needed to deal with complex phenomena;
\item such ``theoretical'' accuracy properties can be achieved by PP estimators
only for very restrictive linearity hypotheses on $f_{o}$. 
\end{itemize}

\section{Hypotheses Falsification Properties of Estimators\label{sec:Hypotheses-Falsification}}

In the previous section, it has been shown that, if the BA conditions
hold true, estimators $\phi\in\Phi\left(D_{o},F^{a},D^{a}\right)$
enjoying relevant theoretical accuracy properties can be derived.
In this section, the problem is discussed if, making use of the available
data $D_{o}$, the BA conditions can be verified, or validated, as
often referred to in the literature. 

The following result holds.
\begin{thm}
\label{thm:Th3}Proving that the BA conditions are verified is not
possible, whatever large the data set $D_{o}$ is.
\end{thm}
\textbf{Proof.} The theorem is proved by showing that it holds for
exact measurements, where $D_{\mathit{ex}}=\left\{ u_{k},y_{k};k=1,\ldots,M\right\} $
and $\mathit{FFS}\left(D_{\mathit{ex}},F^{a}\right)=\left\{ f\in F^{a}:y_{k}=f\left(u_{k}\right),k=1,\ldots,M\right\} $.
\textit{A }\textit{fortiori}, it holds in the case of corrupted measurements.

Let us consider that $\mathit{FFS}\left(D_{\mathit{ex}},F^{a}\right)$
is not empty and bounded. If empty, from Theorem~\ref{thm:Th2} it
follows that no reliable estimator exists. Thus, the theorem is proved
showing that, making use of the information provided by data $D_{\mathit{ex}}$,
the necessary hypothesis $f_{o}\in\overline{FFS}\left(D_{o},F^{a},D^{a}\right)$
cannot be proved. Since $\mathit{FFS}\left(D_{\mathit{ex}},F^{a}\right)$
is bounded, also the set $\overline{FFS}\left(D_{\mathit{ex}},F^{a}\right)$
is bounded and a finite constant $B$ exists such that $\underset{f\in\overline{FFS}\left(D_{\mathit{ex}},F^{a}\right)}{\sup}\left\Vert f\right\Vert _{q}^{U}\leq B$.
Data $D_{\mathit{ex}}$ could have been generated by any function
$f$ belonging to the set $\mathit{FFS}\left(D_{\mathit{ex}}\right)=\left\{ f\in C^{0}:y_{k}=f\left(u_{k}\right),k=1,2,\ldots,M\right\} $.
In the proof of Lemma~\ref{lem:Lemma1} it is shown that set $\mathit{FFS}\left(D_{\mathit{ex}}\right)$
is unbounded, thus certainly containing functions $f^{B}$ such that
$\left\Vert f\right\Vert _{q}^{U}>B$, not belonging to $\overline{FFS}\left(D_{\mathit{ex}},F^{a}\right)$,
and it is not excluded that $f_{o}$ is one of them.\hfill$\blacksquare$

\bigskip{}

From Theorem~\ref{thm:Th3} it follows that, whatever hypotheses
$F^{a},D^{a}$ are considered, no one of the derived theoretical properties
of estimator $\widehat{\phi}\in\Phi\left(D_{o},F^{a},D^{a}\right)$
can be claimed to hold when the estimate $\widehat{f}\left(\widetilde{u}\right)$
is used to approximate $f_{o}\left(\widetilde{u}\right)$ for new
$\widetilde{u}\in U$. For example, let $\phi^{\ast}\in\Phi\left(D_{o},F^{a},D^{a}\right)$
be optimal for $q=\infty$, with ``small'' $E_{\infty}\left(\phi^{\ast}\right)$,
theoretically guaranteeing that the error $\left|f_{o}\left(\widetilde{u}\right)-f^{\ast}\left(\widetilde{u}\right)\right|\leq E_{\infty}\left(\widehat{\phi}\right)$,
for any new $\widetilde{u}\in U$. However, no finite bound on error
$\left|f_{o}\left(\widetilde{u}\right)-f^{\ast}\left(\widetilde{u}\right)\right|$
can be claimed for new measured $\widetilde{u}\in U$, since it cannot
be proved that conditions BA hold true.

\bigskip{}

Since verifying that the conditions required for deriving reliable
estimators \\ $\widehat{\phi}\in\Phi\left(D_{o},F^{a},D^{a}\right)$
is not viable, it is now investigated if it is possible to recognize
hypotheses $F^{a}$, $D^{a}$ that cannot lead to reliable estimators
$\widehat{\phi}\in\Phi\left(D_{o},F^{a},D^{a}\right)$. From Theorem~\ref{thm:Th2}
it follows that if hypotheses $F^{a},D^{a}$ are such that $\mathit{FFS}\left(D_{o},F^{a},D^{a}\right)$
is empty, then no reliable estimator $\widehat{\phi}\in\Phi\left(D_{o},F^{a},D^{a}\right)$
may exist. Such hypotheses are called ``falsified'', according to
the falsification concept introduced in the seminal contributions
of Popper, see, e.g., \cite{Popp35}. Indeed, if $\mathit{FFS}\left(D_{o},F^{a},D^{a}\right)$
is empty, hypotheses $F^{a}$, $D^{a}$ are falsified by data $D_{o}$,
i.e., no functions $f\in F^{a}$ exist that can generate the measured
data $\left\{ \widetilde{u}_{k},\widetilde{y}_{k};k=1,\ldots,M\right\} $,
with disturbances $d_{k}\in D^{a}$, see (\ref{eq:FFS}).

Thus, the falsification concept is formalized as follows.

\textbf{Falsification criterion:}
\begin{enumerate}
\item hypotheses $f_{o}\in F^{a}$\textit{,} $d\in D^{a}$ are falsified
by data $D_{o}$ if $\mathit{FFS}\left(D_{o},F^{a},D^{a}\right)$
is empty
\item hypotheses $f_{o}\in F^{a}$\textit{,} $d\in D^{a}$ are unfalsified
by data $D_{o}$ if $\mathit{FFS}\left(D_{o},F^{a},D^{a}\right)$
is not empty 
\end{enumerate}
For given hypotheses $F^{a}$, $D^{a}$, the falsification concept
may allow to classify them as falsified or unfalsified, provided that
hypotheses $f_{o}\in F^{a}$, $d\in D^{a}$ have the property that
it is possible to prove the $\mathit{FFS}\left(D_{o},F^{a},D^{a}\right)$
emptiness, called ``falsification'' property. This property has
an important role, because recognizing that hypotheses $F^{a}$, $D^{a}$
are falsified by data $D_{o}$ allows to discard hypotheses on function
$f_{o}$ and disturbance $d$ not consistent with the measured data.

Another important aspect of falsification concept is that hypotheses
$F^{a}$, $D^{a}$ unfalsified by data $D_{o}$ may be falsified by
new measurements. Let a new measurement $D^{\mathit{1}}=\left\{ \widetilde{u},\widetilde{y}\right\} $
be collected, and data set $D_{o}^{\mathit{1}}=\left\{ D_{o},D^{\mathit{1}}\right\} $
be considered. Set $\mathit{FFS}\left(D_{o}^{\mathit{1}},F^{a},D^{a}\right)$
may be not empty or empty. If not empty, the hypotheses $F^{a}$,
$D^{a}$ remain classified as unfalsified using data $D_{o}^{\mathit{1}}$.
If empty, the hypotheses $F^{a}$, $D^{a}$, unfalsified using data
$D_{o}$, are falsified by data $D_{o}^{\mathit{1}}$, and new hypotheses
$F_{\mathit{new}}^{a}$, $D_{\mathit{new}}^{a}$ unfalsified by data
$D_{o}^{\mathit{1}}$ are selected.

Falsification certainly happens if $\widetilde{u}\notin U$ and $f_{o}\left(u\right)\notin F^{a}$
for $u\notin U$. Falsification may also happen if $\widetilde{u}\in U$,
but, as shown below, it may also never happen $\forall\widetilde{u}\in U$.

Recognizing that unfalsified hypotheses $F^{a}$, $D^{a}$ are falsified
by new measurements is important in science, since new hypotheses
have to be considered, possibly leading to new scientific discoveries,
as happened on several occasions in the history of science.

Thus, the falsification property of hypotheses is a key distinctive
feature that allows to distinguish between Science and Pseudo-Sciences. 

Below the falsification properties of PP and SM hypotheses are now
discussed.

\subsection{Falsification properties of PP hypotheses\label{subsec:PP-Falsification}}

Any PP hypotheses $f_{o}\in F_{p}^{a},d\in D_{s}^{a}$ are unfalsified
if the support of the probability density functions assumed for disturbance
$d$ is unbounded, since, whatever $F_{p}^{a}$ is considered, $f_{p}\in F_{p}^{a}$
exist such that $\widetilde{y}_{k}=f_{p}\left(\widetilde{u}_{k}\right)+d_{k},k=1,\ldots,M$
for suitable $d_{k}$. However, if $F_{p}^{a}$ has low dimension,
typically measured by $n_{p}$, some $d_{k}$ may be out of the $\alpha$
\%-confidence interval $\left|d\right|\leq\delta\left(\alpha\right)$.
Thus, if the following set is defined $\mathit{FFS}\left(D_{o},F_{p}^{a},D^{a}\right)=\left\{ f\in F_{p}^{a}:\left|\widetilde{y}_{k}-f\left(\widetilde{u}_{k}\right)\right|\leq\delta\left(\alpha\right),k=1,\ldots,M\right\} $,
it is possible to reject \textquotedblleft too poor\textquotedblright{}
parametric families $F_{p}^{a}$, e.g., using Support Vector Regression
(SVR) algorithm to evaluate if $\mathit{FFS}\left(D_{o},F_{p}^{a},D^{a}\right)$
is empty. Note that, from (\ref{eq:measurements}), $d=f_{o}\left(u\right)-f_{o}(u+\mathit{du})-\mathit{dy}$.
Even if the statistical properties of errors $\mathit{du},\mathit{dy}$
can be reliably known from data sheets of the sensors measuring the
variables $u,y$, it may not be easy to evaluate the $\alpha$\%-confidence
interval for given value, but some result can be found in SVR literature,
see, e.g., \cite{Vapn00}.

\subsection{Falsification properties of SM hypotheses\label{subsec:SM-Falsification}}

Let $\overline{f}\left(u\right)$ be the function defined in (\ref{eq:f_upper}).
Necessary and sufficient conditions for falsification of hypotheses
$f_{o}\in F_{\gamma}^{a}$ and $\widetilde{d}\in D_{\varepsilon}^{a}$
are (see \cite[Theorem~1]{MiNo04}):
\begin{enumerate}
\item Sufficient condition: $\overline{f}\left(\widetilde{u}_{k}\right)-\widetilde{y}_{k}+\varepsilon<0,\quad k=1,2,\ldots,M$
\item Necessary condition: $\overline{f}\left(\widetilde{u}_{k}\right)-\widetilde{y}_{k}+\varepsilon\leq0,\quad k=1,2,\cdots,M$
\end{enumerate}
The function $\gamma^{\ast}\left(\varepsilon\right)=\underset{\varepsilon:\overline{f}\left(\widetilde{u}_{k}\right)\leq\widetilde{y}_{k}-\varepsilon,k=1,2,\ldots,M}{\mathit{\inf}}\gamma\:$
individuates a curve, called the falsification curve, that separates
the space $(\gamma,\varepsilon)$ in two regions: 
\begin{itemize}
\item Falsified Region: the set of values $(\gamma,\varepsilon)$ for which
the hypotheses $f_{o}\in F_{\gamma}^{a}$ and $\widetilde{d}\in D_{\varepsilon}^{a}$
are falsified;
\item Unfalsified Region: the set of values $(\gamma,\varepsilon)$ for
which the hypotheses $f_{o}\in F_{\gamma}^{a}$ and $\widetilde{d}\in D_{\varepsilon}^{a}$
are unfalsified.
\end{itemize}
\begin{figure}[h]
\centering
\centering{}\includegraphics[width=10cm,height=3.6654in]{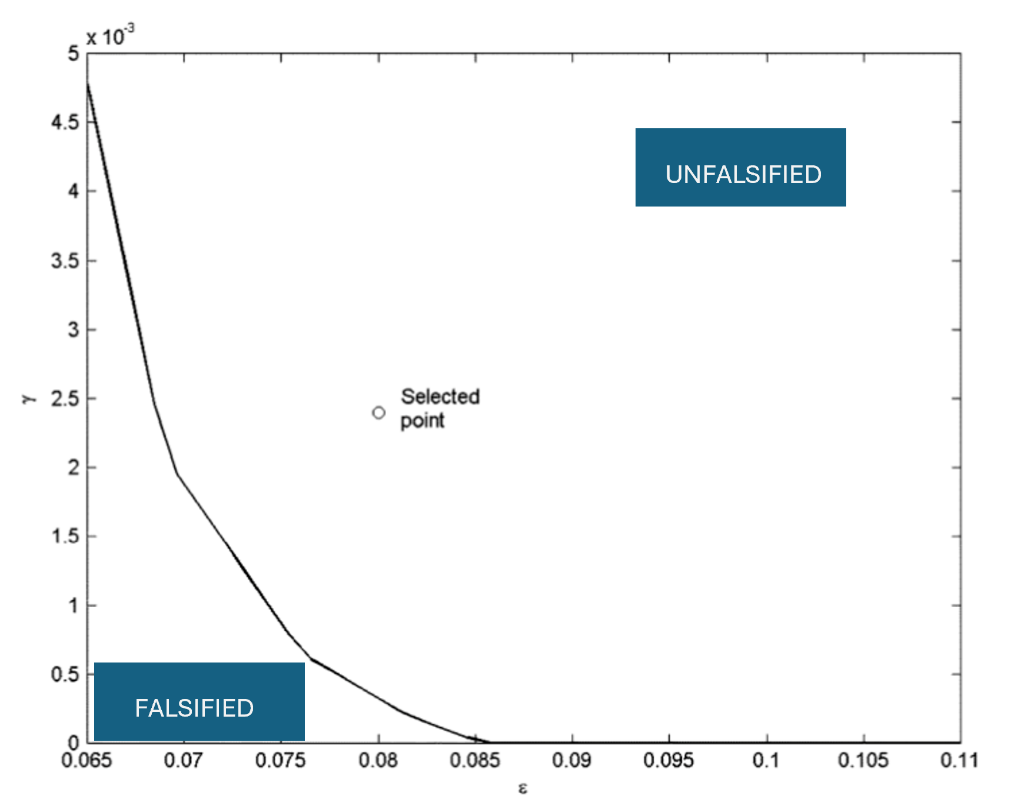}\caption{Falsification curve, Falsified and Unfalsified Regions}
\label{Falsification_curve}
\end{figure}

As previously discussed, recognizing if hypotheses $F_{\gamma}^{a}$,
$D_{\varepsilon}^{a}$, unfalsified by data $D_{o}$, are falsified
by new measurements $D^{\mathit{1}}=\left\{ \widetilde{u}^{i},\widetilde{y}^{i},i=1,2,\ldots\right\} $
is important. An efficient recursive method can be obtained as follows.
Let $\gamma,\varepsilon$ values belong to the Unfalsified Region,
i.e., $\mathit{FFS}\left(D_{o},F_{\gamma}^{a},D_{\varepsilon}^{a}\right)$
is not empty. Let $i_{1}$ be the first index such that $\overline{f}\left(\widetilde{u}^{i_{1}}\right)-\widetilde{y}^{i_{1}}+\varepsilon>0$
and $D_{o}^{1}=\left\{ D_{o},\widetilde{u}^{i_{1}},\widetilde{y}^{i_{1}}\right\} $.
From the Necessary condition above, $\mathit{FFS}\left(D_{o}^{1},F^{a},D^{a}\right)$
is empty and then hypotheses $F_{\gamma}^{a}$, $D_{\varepsilon}^{a}$
are falsified by the measurements $\widetilde{u}^{i_{1}}$, $\widetilde{y}^{i_{1}}$.
Then, new unfalsified hypotheses $F_{\gamma_{1}}^{a}$, $D_{\varepsilon_{1}}^{a}$
can be obtained by selecting $\gamma_{1}>\gamma$, $\varepsilon_{1}>\varepsilon$
belonging to the Unfalsified Region computed using the data set $D_{o}^{1}=\left\{ D_{o},\widetilde{u}^{i_{1}},\widetilde{y}^{i_{1}}\right\} $.
Let $D_{o}^{j}=\left\{ D_{o},\widetilde{u}^{i_{1}},\widetilde{y}^{i_{1}},\ldots,\widetilde{u}^{i_{j}},\widetilde{y}^{i_{j}}\right\} $
the data set after $j$ such falsification events, $U^{j}$ the smallest
box containing $D_{o}^{j}$ and $\gamma_{j},\varepsilon_{j}$ the
last $\gamma,\varepsilon$ values selected. The question is now discussed
if it can be conjectured that for some $j=n$ it may happen that no
new data will falsify hypotheses $F_{\gamma_{n}}^{a}$, $D_{\varepsilon_{n}}^{a}$.
Indeed, at each falsification event $j$, $\gamma_{j}>\gamma_{j-1},\varepsilon_{j}>\varepsilon_{j-1}$,
$\mathit{FFS}\left(D_{o}^{j},F_{\gamma_{j}}^{a},D_{\varepsilon_{j}}^{a}\right)\supset$$\mathit{FFS}\left(D_{o}^{j},F_{\gamma_{j-1}}^{a},D_{\varepsilon_{j-1}}^{a}\right)$.
Thus, if $\gamma_{n},\varepsilon_{n}$ values are sufficiently large,
then it may happen that $f_{o}\in\overline{\mathit{FFS}}\left(D_{o}^{n},F_{\gamma_{n}}^{a},D_{\varepsilon_{n}}^{a}\right)$.
For example, this happens if $f_{o}\in C^{k},k\geq1$, being $C^{k}$
the class of all functions that are differentiable $k$ times and
whose $k$-th order derivatives are continuous, since values $\gamma_{n}\geq\gamma^{o}=\underset{u\in U}{\mathit{max}}\left\Vert f_{o}^{\prime}\left(u\right)\right\Vert _{2}$
and $\varepsilon_{n}$ exist in the Unfalsified Region computed using
the data set $D_{o}^{j}$, such that actually $f_{o}\in\mathit{FFS}\left(D_{o}^{n},F_{\gamma_{n}}^{a},D_{\varepsilon_{n}}^{a}\right)$.
For such hypotheses $F_{\gamma_{n}}^{a},D_{\varepsilon_{n}}^{a}$,
both conditions of Theorem~\ref{thm:Th2} hold, and consequently
the estimator $\phi^{c}\in\Phi\left(D_{o}^{n},F_{\gamma_{n}}^{a},D_{\varepsilon_{n}}^{a}\right)$
is optimal and no new data $\widetilde{u}^{i}\in U^{n}$ will falsify
hypotheses $F_{\gamma_{n}}^{a},D_{\varepsilon_{n}}^{a}$.

\section{Parametric Set Membership (PSM) estimators\label{sec:Parametric-Set-Membership}}

As discussed in the previous sections, PP estimators $\phi_{\widehat{p}}\in\Phi\left(D_{o},F_{p}^{a},D_{s}^{a}\right)$,
where $F_{p}^{a}$ is a set of complex nonlinear functions, have less
strong accuracy and falsification properties than SM estimators. However,
estimated $f_{\widehat{p}}$ may provide useful information on $f_{o}$,
especially if the parametric family of functions $F_{p}^{a}$ is obtained
from physical laws based on previous measurements on the phenomenon
under investigation. In this section, it is shown how a new class
of estimators can be obtained, called PSM estimators, able to integrate
the information provided by $f_{\widehat{p}}$ with a suitably designed
SM estimator.

\bigskip{}

The unknown residual function $f_{\Delta}=f_{o}-f_{\widehat{p}}$
is estimated making use of the data set $D_{\Delta}^{o}=\left\{ \widetilde{u}_{k},\Delta\widetilde{y}_{k},k=1,\ldots,M\right\} $,
where $\Delta\widetilde{y}_{k}=\widetilde{y}_{k}-f_{\widehat{p}}\left(\widetilde{u}_{k}\right)$
and considering the SM hypotheses $f_{\Delta}\in F_{\gamma_{\Delta}}^{a}$
and $\left|\widetilde{y}_{k}-f\left(\widetilde{u}_{k}\right)\right|\leq\varepsilon$,
$\left|\Delta\widetilde{y}_{k}-f_{\Delta}\left(\widetilde{u}_{k}\right)\right|<\varepsilon_{\Delta}$,
with $(\gamma_{\Delta},\varepsilon_{\Delta})$ values belonging to
the Unfalsified Region. A central SM estimator $\phi_{\Delta}^{c}$
is derived, giving a theoretical optimal estimate $f_{\Delta}^{c}$
of the unknown function $f_{\Delta}=f_{o}-f_{\widehat{p}}$. The PSM
estimator $\phi_{P}^{c}$ is obtained taking $f_{P}^{c}=f_{\widehat{p}}+f_{\Delta}^{c}$
as estimate of $f_{o}$.

The rationale behind the introduction of PSM estimators is as follows.
If the assumption $\left\Vert f_{o}^{\prime}\left(u\right)-f_{\widehat{p}}^{\prime}\left(u\right)\right\Vert _{2}\leq\gamma_{\Delta}$
is adopted on $f_{\Delta}$, then $\left\Vert f_{o}^{\prime}\left(u\right)\right\Vert _{2}\leq\left\Vert f_{\widehat{p}}^{\prime}\left(u\right)\right\Vert _{2}+\gamma_{\Delta}$
holds for $f_{o}$. Thus, the PSM estimator $\phi_{P}^{c}$ is derived
assuming on $f_{o}\left(u\right)$ a bound on its gradient modulus
$\left\Vert f_{o}^{\prime}\left(u\right)\right\Vert _{2}$ that is
variable with its argument $u$, with the minimal value $\gamma_{\Delta}$
for values of $u$ such that $\left\Vert f_{\widehat{p}}^{\prime}\left(u\right)\right\Vert _{2}\cong0$.
Thus, the PSM estimator $\phi_{P}^{c},$ integrating the information
provided by the physical estimator $\phi_{\widehat{p}}$, may be more
accurate than an estimator $\phi^{c}$ based only on SM assumption
$f_{o}\in F_{\gamma}^{a}$, where a constant value $\gamma$ is assumed
such that $\left\Vert f^{\prime}\left(u\right)\right\Vert _{2}\leq\gamma,\forall u\in U$.

The following theoretical results hold for the SM estimator $\phi_{\Delta}^{c}$,
that will be used for deriving theoretical properties of the PSM estimator
$\phi_{P}^{c}$.

The unknown residual function $f_{\Delta}$ is tightly bounded as:
\begin{equation}
\underline{f_{\Delta}}\left(\widetilde{u}\right)\leq f_{\Delta}\left(\widetilde{u}\right)\leq\overline{f_{\Delta}}\left(\widetilde{u}\right),\quad\forall\widetilde{u}\in U\label{eq:f_Delta_bounds}
\end{equation}
where $\underline{f_{\Delta}}\left(u\right)=\underset{k=1,\ldots,M}{\max}\left[\left(\Delta\widetilde{y}_{k}-\varepsilon\right)-\gamma_{\Delta}\cdot\left\Vert u-\widetilde{u}_{k}\right\Vert _{2}\right]$
and \\ $\overline{f_{\Delta}}\left(u\right)=\underset{k=1,\ldots,M}{\min}\left[\left(\Delta\widetilde{y}_{k}+\varepsilon\right)+\gamma_{\Delta}\cdot\left\Vert u-\widetilde{u}_{k}\right\Vert _{2}\right]$.
The estimate $f_{\Delta}^{c}\left(u\right)=\left[\underline{f_{\Delta}}\left(u\right)+\overline{f_{\Delta}}\left(u\right)\right]/2$
is an optimal approximation of the residual function $f_{\Delta}=f_{o}-f_{\widehat{p}}$
for any $L_{q}\left(U\right)$ norm, $q=1,\ldots,\infty.$

The error of the SM estimator $\phi_{\Delta}^{c}$ is:
\begin{equation}
E^{a}\left(\phi_{\Delta}^{c}\right)=\underset{f_{\Delta}\in FFS_{\Delta}}{\sup}\left\Vert f_{\Delta}-f_{\Delta}^{c}\right\Vert _{q}^{U}=\left\Vert \underline{f_{\Delta}}-\overline{f_{\Delta}}\right\Vert _{q}^{U}/2\label{eq:SM_estimator_error}
\end{equation}
From (\ref{eq:f_lower}), (\ref{eq:f_upper}), (\ref{eq:f_Delta_bounds}),
(\ref{eq:SM_estimator_error}) and the fact that $f_{P}^{c}=f_{\widehat{p}}+f_{\Delta}^{c}$,
the following theoretical results of the PSM estimator $\phi_{P}^{c}$
hold.

The value $f_{o}\left(u\right)$ is tightly bounded as:
\begin{equation}
f_{\widehat{p}}\left(u\right)+\underline{f_{\Delta}}\left(u\right)\leq f_{o}\left(u\right)\leq f_{\widehat{p}}\left(u\right)+\overline{f_{\Delta}}\left(u\right),\quad\forall u\in U\label{eq:f_o_bounds}
\end{equation}
The PSM estimator $\phi_{P}^{c}$ is optimal in estimating $f_{o}$
for any $L_{q}\left(U\right)$ norm, $q=1,\ldots,\infty$, and its
error is:
\begin{equation}
E^{a}\left(\phi_{P}^{c}\right)=\sup_{f\in\mathit{FFS}}\left\Vert f-f_{P}^{c}\right\Vert _{q}^{U}=\left\Vert \underline{f_{\Delta}}-\overline{f_{\Delta}}\right\Vert _{q}^{U}/2\label{eq:PSM_estimator_error}
\end{equation}
The PSM estimator $\phi_{P}^{c}$ is optimal in estimating $f_{o}\left(u\right)$
for any $u\in U$, and its error is:
\begin{equation}
E^{a}\left(\phi_{P}^{c},u\right)=\left|f_{\widehat{p}}+\overline{f_{\Delta}}\left(u\right)-f_{P}^{c}\left(u\right)\right|,\quad\forall u\in U\label{eq:PSM_estimator_error_u}
\end{equation}
A falsification test of PSM assumption can be obtained by applying
the falsification test of SM hypotheses $f_{\Delta}\in F_{\gamma_{\Delta}}^{a}$
and $\left|\widetilde{d}\right|\leq\varepsilon$.

The PSM estimator $\phi_{P}^{c}$ not only may be more accurate than
an estimator $\phi^{c}$ based only on SM assumption, as previously
discussed, but may also give accuracy improvements over the physical
estimator $\phi_{\widehat{p}}$ in the case that the term $f_{\Delta}^{c}$
is not negligible with respect to $f_{\widehat{p}}$, indicating that
the assumed parametric set of functions $F_{p}^{a}$ and/or the statistical
hypotheses $D_{s}^{a}$ used for the derivation of $f_{\widehat{p}}$
are falsified. Even in cases where the correction term $f_{\Delta}^{c}$
is negligible, i.e., the two estimators derive the same approximating
functions $f_{P}^{c}\cong f_{\widehat{p}}$, the PSM estimator $\phi_{P}^{c}$
allows to derive the theoretical results (\ref{eq:f_o_bounds}), (\ref{eq:PSM_estimator_error}),
(\ref{eq:PSM_estimator_error_u}) and to perform falsification analysis.

\bibliographystyle{ieeetr}
\bibliography{mybiblio_new}

\end{document}